\documentclass[reprint, amsmath,amssymb,aps,prb,floatfix,]{revtex4-2}
\usepackage{graphicx}
\usepackage{dcolumn}
\usepackage{bm}
\begin{document}

\preprint{APS/123-QED}

\title{Shear based gap control in 2D photonic quasicrystals of dielectric cylinders}

\author{A. Andueza}

\email{angel.andueza@unavarra.es}
\author{Joaquin Sevilla}

\affiliation{Smart Cities Institute (SCI), Universidad Pública de Navarra, 31006 Pamplona, Spain}%

\author{J. Pérez-Conde}

\affiliation{ Institute for Advanced Materials (INAMAT), Universidad Pública de Navarra, Campus de Arrosadia, 31006 Pamplona, Spain}
\author{K. Wang}

\affiliation{Laboratoire de Physique des Solides, CNRS, Université Paris-Saclay, 91405 Orsay, France  }

\date{\today}

\begin{abstract}

2D dielectric photonic quasicrystals can be designed to show isotropic band gaps. The system here studied is a quasiperiodic lattice made of silicon dielectric cylinders ($\varepsilon$ = 12) arranged as periodic unit cell based on a decagonal approximant of a quasiperiodic Penrose lattice. We analyze the bulk properties of the resulting lattice as well as the bright states excited in the gap which correspond to localized resonances of the electromagnetic field in specific cylinder clusters of the lattice. Then we introduce a controlled shear deformation $\gamma$ which breaks the decagonal symmetry and evaluate the width reduction of the gap together with the evolution of the resonances, for all shear values compatible with physical constraints (cylinder collision). The gap is reduced up to a $18.5 \%$ while different states change their frequency in different ways. Realistic analysis about the actual transmission of the electromagnetic radiation, often missing in the literature, have been performed for a finite “slice” of the proposed quasicrystals  structure. Two calculation procedures based on MIT Photonic Bands (MPB) and Finite Integration Technique (FIT) are used for the bulk and the finite structures finding an extremely good agreement among them.

\end{abstract}


\maketitle

\section{\label{sec:intro}Introduction }

Photonic band gap (PBG) materials have been investigated for more than 
three decades since the germinal paper of Ohtaka \cite{ohtaka1979,notomi2010}. One of the main
objectives of the past PBG effort was the search of metamaterials with new
optical related properties like isotropic band gap \cite{man2013a}, 
zero-refractive index \cite{boriskina2015}, slow light manipulation 
\cite{notomi2010}, etc. These PBG allow to create waveguides 
\cite{priyarose2012}, sensors \cite{andueza2016a}, collimators
\cite{matthews2009} or to improve the existent solar cells \cite{cornago2015}, 
etc. 

Some realistic PBG materials were early proposed in three 
\cite{yablonovitch1989} and two dimensions \cite{villeneuve1992,meade1992} which were based on periodic lattices with complete but anisotropic band gaps (see for instance Ref. \citenum{joannopoulos2008},
specially chapter 5 for 2D PC's). Isotropic band gaps are preferred, though, 
for some applications such as wavelength selective mirror or filters
\cite{edagawa2014} slow light technology \cite{baba2008}, filters 
\cite{ali2011}, 
collimators \cite{matthews2009} or band filters \cite{ali2011}.

On the other hand, all-dielectric photonic crystals avoid metals and therefore can be fabricated with only dielectric components \cite{edagawa2016,staude2019}. Meta-optics based in Mie-type resonances in all-dielectric nanostructures is an emerging field \cite{staude2019}.

The creation of photonic materials with complete band gap is also important to control spontaneous emission and Purcell effect \cite{staude2019}.
All-dielectric photonic quasicrystals can be also thought as bandgap systems 
\cite{wang2007,wang2010,vardeny2013} or as zero-refractive index homogeneous materials \cite{boriskina2015}.

Photonic quasicrystals (PQC) have been proposed and built with dielectric 
cylinders \cite{zoorob2000,wang2003a}. Some of these structures presented an almost isotropic band gap which is one of the required features to build most of the applications. Since then many QC structures based on dielectric cylinders
have been investigated \cite{romero-vivas2005,wang2006,wang2007,priyarose2012,vardeny2013}. In these works high permittivity contrast (cylinder vs surrounding material) was needed to get a complete and sizable band gap as it was in periodic two dimensional PC case \cite{notomi2010}. Recently, low refractive index contrast \(1.6:1\) PBG's based on hyperuniform 2D structures have been also investigated \cite{man2013a} and demonstrated the possibility to build PBG with low/medium refractive indexes.

An important issue with PGB and PQC in particular, is to create a device with a gap which could be also modified in controlled way. We propose here a way to
create such structures, which show an almost isotropic gap, that can be also controlled with only geometrical parameters. We study an arrangement of silicon cylinders ($\varepsilon$ = 12) arranged in a (regular) lattice of Penrose quasicrystal approximants, a generalization of previously studied in Ref. \citenum{wang2010}.

In previous studies it has been shown the onset of states in the gap due
Mie-like localized resonances \cite{wang2006,wang2007,wang2010}. These 
resonances have been previously reported within a band structure context  without description
of their the actual transmission properties. Here we calculate their transmission output and their evolution with the shear of the underlying
lattice together with the band structure of the bulk lattice counterpart. We
also analyze the group velocity of these resonances, which
is very small \cite{wang2007}, an important feature for applications where slow light
technology is needed \cite{notomi2010}.

Finally, the simulations have been realized assuming that the cylinders are surrounded by air, although the results should remain valid for cylinders in a dielectric matrix if the permittivity contrast is kept.

\section{\label{sec:system}System under study and calculation procedures}

We want to study, in one hand, the effect of symmetry braking by shear in the gap of photonic quasicrystals and, on the other, how this variation affects light transmission through the material. The gap calculations are performed on an infinite periodical system, while the transmission data must be obtained from a finite sample in the incidence direction. In the following subsections we describe how both systems are built (\ref{subsec:appr}), how shear is applied (\ref{subsec:shear}) and the numerical methods used for the calculations (\ref{subsec:num}), MIT Photonic Band (MPB) and CST Microwave Studio (CST) for photonic band gap and transmission spectra calculations, respectively. 

\begin{figure}[htbp]
	\centering
	\includegraphics[width=0.46\textwidth]{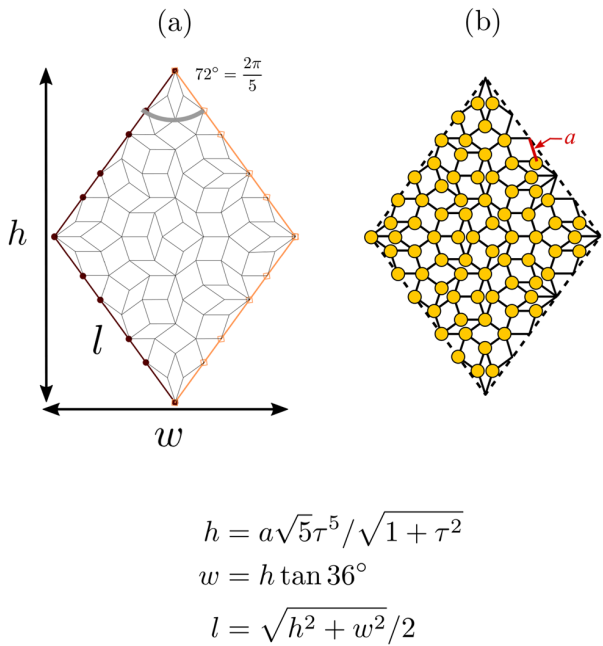}
	\caption{\label{fig:app_deca} (a) Decagonal approximant in the original QC. The unit cell is delimited by dashed lines is shown as well as the 
	(b) dielectric cylinders (yellow circles) at the lattice nodes.}
\end{figure}

\subsection{\label{subsec:appr}Quasicrystal approximants and Bravais lattices}

The bulk quasicrystal photonic lattices are not 
suitable to build a physical realization nor to perform transmission calculations on them. In this work we define approximants of the parent QC as the unit cell of an otherwise periodic lattice as was first proposed in Ref.
\citenum{wang2003a}. The approximant-based lattice reproduces most of the 
properties of the parent QC lattice, such as  tenfold rotational symmetries, long-range decagonal bond-orientational
symmetries and photonic band gap. In addition, it allows to use the standard computational methods developed for periodic photonic crystals (see Fig. \ref{fig:app_deca}). 

When transmission computations are concerned, we
need a lattice which is finite in the propagation direction to obtain a sizable
output. As the signal intensity decreases with the lattice width, 
we must therefore arrive to a compromise between
contrast between gap and resonant states intensity, which needs a big enough crystal, and measurable transmission. 
On the other hand, we know that at least three wavelength width in the propagation direction is necessary to get an adequate contrast between the in-gap states and the gap background in previous work \cite{andueza2016a}.

The decagonal approximant produces an oblique \(2\pi/5\) rhombus-shaped Bravais
lattice (see Fig. \ref{fig:app_deca}) and needs a more complex procedure for the transmission simulations. CST is based on a Finite Integration Technique (FIT) \cite{weiland2003} and allows only to define orthogonal bases, so we had to adapt the decagonal oblique 
approximant to this constraint. 

\begin{figure}[htbp]
	\centering
	\includegraphics[width=0.46\textwidth]{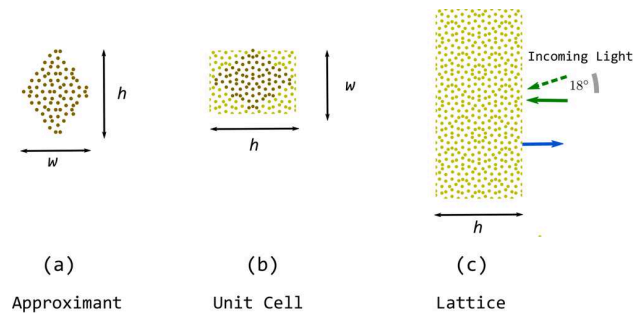}
	\caption{\label{fig:actual_lattice}  Approximant and unit cells for 
	decagonal approximant  (brown) and computational sized \(h\times w\) cell 
	used in transmission CST calculations. The specific values of $ h, w, r $ are also given, where $ r =0.25\,\mathrm{a} $ is the cylinder radius. Blue arrows are the normal to the QC surface and green arrows indicate the range of possible different incoming light directions.}
\end{figure}

The procedure to get an unit cell adapted to a rectangular basis is as follows. We added to the original rhombus four triangles in order to complete a rectangle as shown in Fig. \ref{fig:actual_lattice} (b). From this cell
we can easily create an ordinary rectangular Bravais lattice where each lattice
point contains now \(152\) cylinders, as shown in Fig. \ref{fig:actual_lattice}
(b). The original approximant width was \(w=a \sqrt{5}\tau^3\) and height \(h=a
\sqrt{5}\tau^5/\sqrt{1+\tau^2}\), where \(a\) is the edge length of the tiles that build up the decagonal approximant employed and $ \tau $  is the golden number. The rod radius considered is $ r=0.25\,\mathrm{a} $.

The new rectangular unit size is correspondingly \(w\times h\)  as shown in Fig. \ref{fig:actual_lattice} (b). The size of the lattice in the propagation direction, \(h\), is wide enough to meet the previous requirements of contrast between gap and in-gap states and, on the other hand, it is thin enough to allow transmission measurements \cite{andueza2016a}. The band structure, though, is calculated as a whole 2D lattice with the MPB package \cite{johnson2001} with the rhombus unit cell containing \(76\) rods as shown in Fig. \ref{fig:actual_lattice}.  The final approximant shapes and sizes used with MPB technique are depicted in Fig.\ref{fig:actual_lattice}(c).

\begin{figure}[b]
	\centering
	\includegraphics[width=0.46\textwidth]{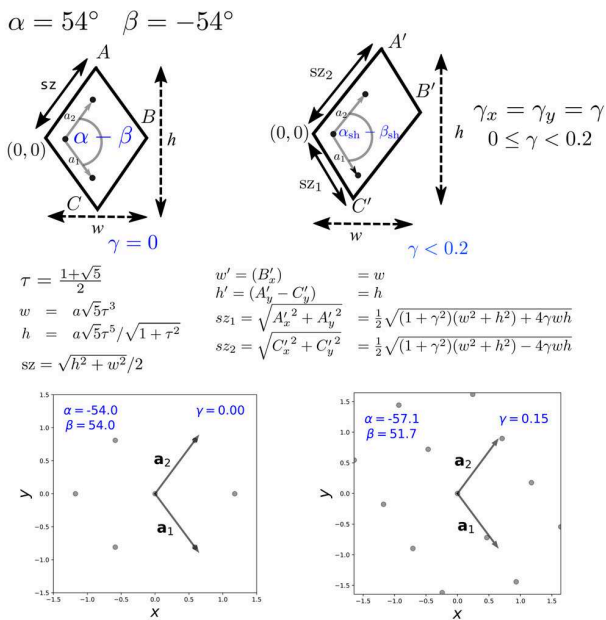}	
	\caption{\label{fig:app_brav_shear} Unit cells of decagonal approximants in the original QC and in the sheared case used with the MPB technique (top). Below we show the Bravais lattice where each point represent an approximant for two values of shear $ \gamma =0.0 $ and $ \gamma =0.15 $. The angles $ \alpha $ and $ \beta $ are measured from the $ y$-axis to the primitive vectors $ \mathbf{a}_{1} $ and $ \mathbf{a}_{2} $ respectively.}
\end{figure}

\subsection{\label{subsec:shear}Broken rotational symmetry by shear}

We investigate the QC behavior as we shear the lattice and, therefore,when the local decagonal rotational symmetry is broken, leaving unmodified the circular shape of the dielectric rods. Shear is one the three possible transformations that can be applied to a lattice. A generic strain or deformation, \(\mathbf{D}\), in two dimensions can be implemented with a deformation applied to every \((x,y)\) point, so that the new transformed point \((x^{\prime},y^{\prime})\) can be written as \cite{ogden1997},

\begin{equation}
\label{eq:deformation}
\begin{pmatrix}
x^{\prime}  \\
y^{\prime}  \\
\end{pmatrix}=
\mathbf{D} \cdot
\begin{pmatrix}
x \\
y \\
\end{pmatrix}=
\begin{pmatrix}
a & b \\
c & d \\
\end{pmatrix} \cdot
\begin{pmatrix}
x \\
y \\
\end{pmatrix}=
\begin{pmatrix}
ax + by \\
cx +dy  
\end{pmatrix}
\end{equation}

Any lattice deformation , \(\mathbf{D}\), can be decomposed in a dilatation, a rotation and a shear. We are interested in deformations that can modify the band
structure or the lattice transmission spectrum. Also, although global dilatation
do produce a shift of the whole spectra and it could be thought as a modification, this behavior can be also seen as a simple scaling of the lengths and frequencies in the Maxwell equations and no symmetry change is implied. We therefore restrict the possible deformations to those of pure shear which break the rotational n-fold symmetry of the QC. The shear transformations , \(\mathbf{S}\), can be described by the matrix,

\begin{equation}
\mathbf{S}=
\begin{pmatrix}
1 & \gamma_x \\
\gamma_y & 1 \\
\end{pmatrix},
\end{equation}
where \(\gamma_{x}, \gamma_{y}\) are two real numbers which reflect the amount of the shear in the \(x,y\) directions respectively. The transformed points
\((x^{\prime},y^{\prime})\) from \((x,y)\) are given from 
equation \ref{eq:deformation},

\begin{equation}
\label{eq:shear}
\begin{pmatrix}
x^{\prime} \\
y^{\prime} \\
\end{pmatrix}
=
\begin{pmatrix}
1 & \gamma_x \\
\gamma_y & 1 \\
\end{pmatrix}
\cdot
\begin{pmatrix}
x \\
y \\
\end{pmatrix}
=
\begin{pmatrix}
x + y\gamma_x\\
y + x\gamma_y \
\end{pmatrix}
\end{equation}

From now on, we study shear transformations with the same deformation in both
axes, \(\gamma_{x}=\gamma_{y}=\gamma\), so that we only need $\gamma$ parameter to characterize a given shear. In Fig. \ref{fig:app_brav_shear} the 
effects of the shear for the decagonal approximant are shown, together with the corresponding original and sheared Bravais lattices.

\begin{figure}[htbp]
	\centering
	\includegraphics[width=0.46\textwidth]{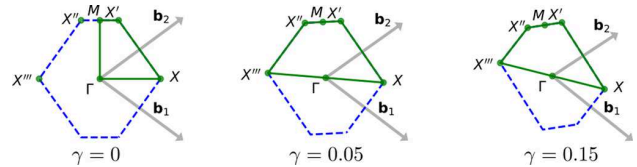}
	\caption{\label{fig:bz_decagonal} Brillouin zone of the approximant-based Bravais lattice  (dashed blue line) together with the irreducible BZ (green) for three values of $\gamma =0, 0.05$ and $0.15$. The irreducible BZ is a fourth of the BZ for $\gamma=0$ but when the rotational  symmetry is broke, $\gamma\neq 0$ is half of of BZ. The  the primitive vectors of the reciprocal lattice,  $\mathbf{b}_1$ and $\mathbf{b}_2$, are also shown.}
\end{figure}

As the shear is applied to the whole quasicrystal the approximants are transformed and the corresponding Bravais lattices are also modified as was shown in Fig. \ref{fig:app_brav_shear}. We need to calculate the new sheared Brillouin zones (BZ), their irreducible part and the special symmetry points to compute the bands at
these special values. In Fig.  \ref{fig:bz_decagonal} we show the original oblique Brillouin zone from decagonal approximant at $\gamma=0$. We also illustrate the $\gamma\neq 0$ instance with two values of $\gamma$, 0.05 and 0.15, respectively. The loss of symmetry in these cases is clearly seen as the irreducible BZ of the sheared lattice is half of the whole BZ, whereas the original unsheared irreducible BZ's for the decagonal lattice was a fourth of the whole BZ. This is the expected behavior for the irreducible BZ: its size is always the \(1/g\mathrm{-th}\) part of the first Brillouin zone (see Ref. \citenum{joshi1997} p. 281). Here \(g\) is the number of symmetry operations of the original point group: $C_{2v}, \,g=4$ in the not sheared oblique  periodic lattice and $C_1, \,g=2$ in the sheared oblique case.

\subsection{\label{subsec:num}Numerical methods}

We compute the periodic approximant-based lattices band by means of
preconditioned conjugate-gradient minimization of the block Rayleigh quotient in
a plane-wave basis, using MIT Photonic Bands package \cite{johnson2001}.

The MPB software allows to inspect in detail the band structure, gap and resonant in-gap states due to light localization. Those quantities describe a periodic and therefore infinite lattice, however, in actual applications, we need to address the experimental accessibility of the photonic lattice, in particular the transmission properties in a finite sample. We investigate this issue by means of Finite  Integration Technique \cite{clemens2001} with the CST MICROWAVE STUDIO\(^{\mathrm{TM}}\), a commercial code. The use of both methods allows, additionally, to cross check the results.

The numerical simulations for a finite version are performed on a two dimensional lattice of dielectric cylinders infinite in the y direction and assuming infinite length cylinders in the  $x$ direction (infinite length rods). The incoming radiation travels in the z direction as Transverse Electric and Magnetic (TEM) mode with the electric field along the x-axis and the magnetic field along the y-axis. The incoming wave presents an incidence angle $\varphi$ of \(18^{\circ}\) to the external surface of the structure as is indicated by the dashed green arrow in Fig. \ref{fig:actual_lattice}(c). A frequency solver tool with a tetrahedral adaptive mesh refinement was used to calculate the power balance of the PQC in CST. The symmetry of the finite PQC and the orientation of the electromagnetic field allow us to simplify the problem restricting the calculation to a rectangular unit cell. CST solves the Maxwell equations and analyzes the weight of each diffraction order separately.  A summation of the power density in all diffraction orders provides the total reflectance, R, and transmittance, T. As the refractive index employed for the calculations is real, the absorptivity of the QC is not considered into the power balance.

We calculate the transmission spectra  with the FIT method for different lattice widths in the $ z $ direction, starting with the largest case ($ h $ wide) as depicted in Fig. \ref{fig:actual_lattice}(b). Then, row after row, the cylinders are removed  and the resulting transmission obtained for each width value. A total of 15 configurations are obtained in the process. The gap and resonances frequencies are then obtained from the analysis of the resulting spectra. The bulk counterpart, as a periodic lattice of decagonal approximants is shown in  Fig. \ref{fig:actual_lattice}(c) for the original (unsheared) and sheared lattices, is analyzed with the MPB package. We obtain the first $ 110 $ bands, which are enough to capture the whole gap frequency range as well as the resonant states in the gap interval.

\section{Results and discussion}

\begin{figure*}
	\centering
	\includegraphics[width=1\textwidth]{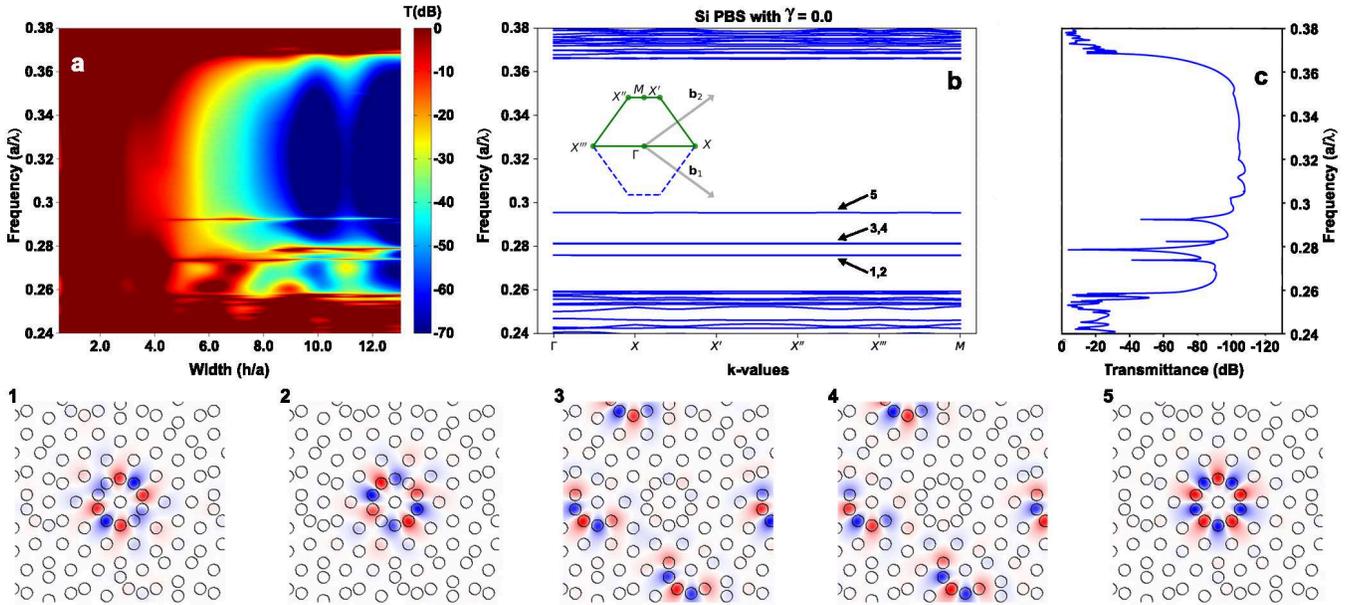}
	\caption{\label{fig:bands_00}Original lattice, $\gamma=0$. (a) Transmission spectra dependence on the normalized width, $ h/a $ (top left color map). (b) Band of the bulk lattice in the same frequency interval, around the gap (top center). The five resonances in the gap are labeled and their field distribution show below. (c) The transmission spectrum for the widest lattice is also shown $ h/a=13$ (top right).}
\end{figure*}

\begin{figure*}[htbp]
	\centering
	\includegraphics[width=1\textwidth]{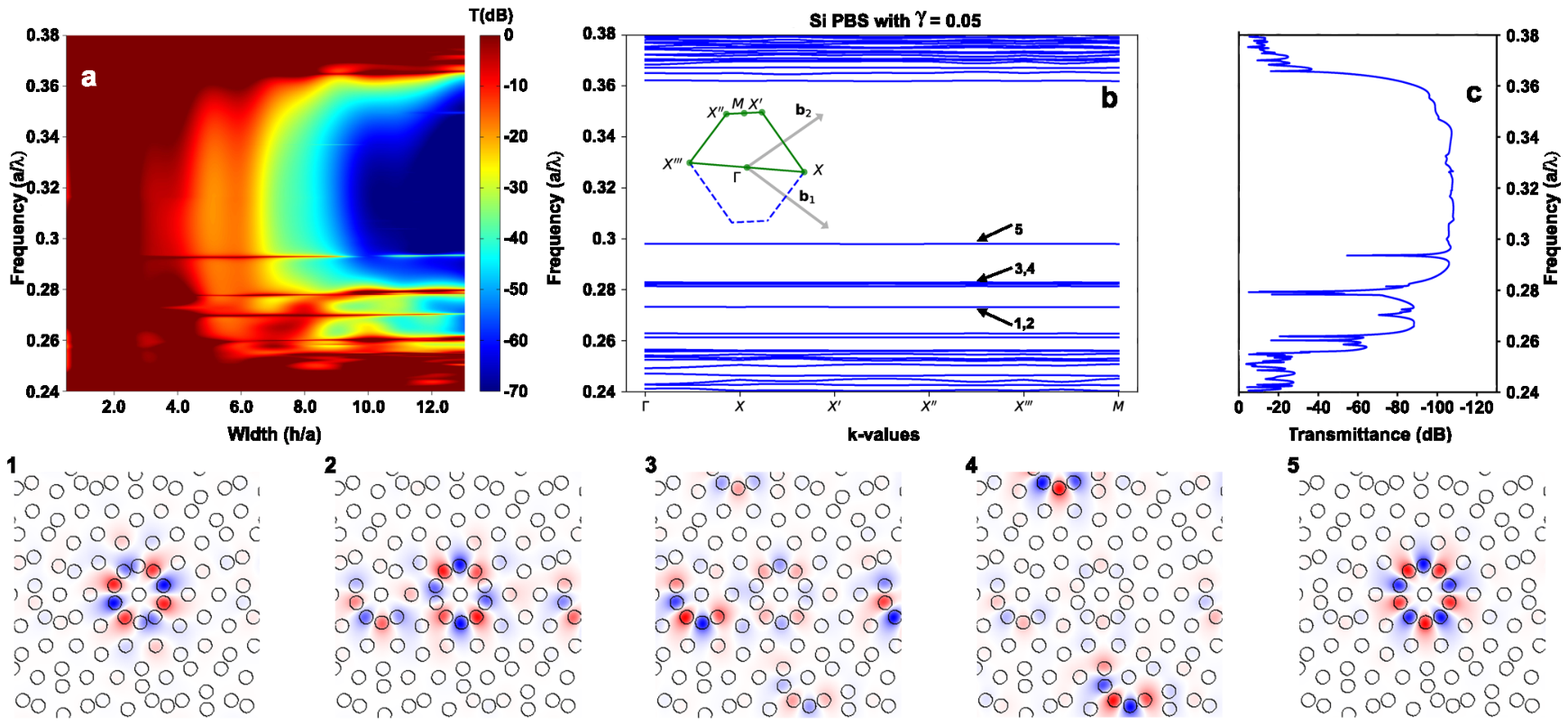}
	\caption{\label{fig:bands_005} Sheared lattice $\gamma=0.05$. The same quantities as in Fig. \ref{fig:bands_00}.}
\end{figure*}

\begin{figure*}[htbp]
	\centering
	\includegraphics[width=1\textwidth]{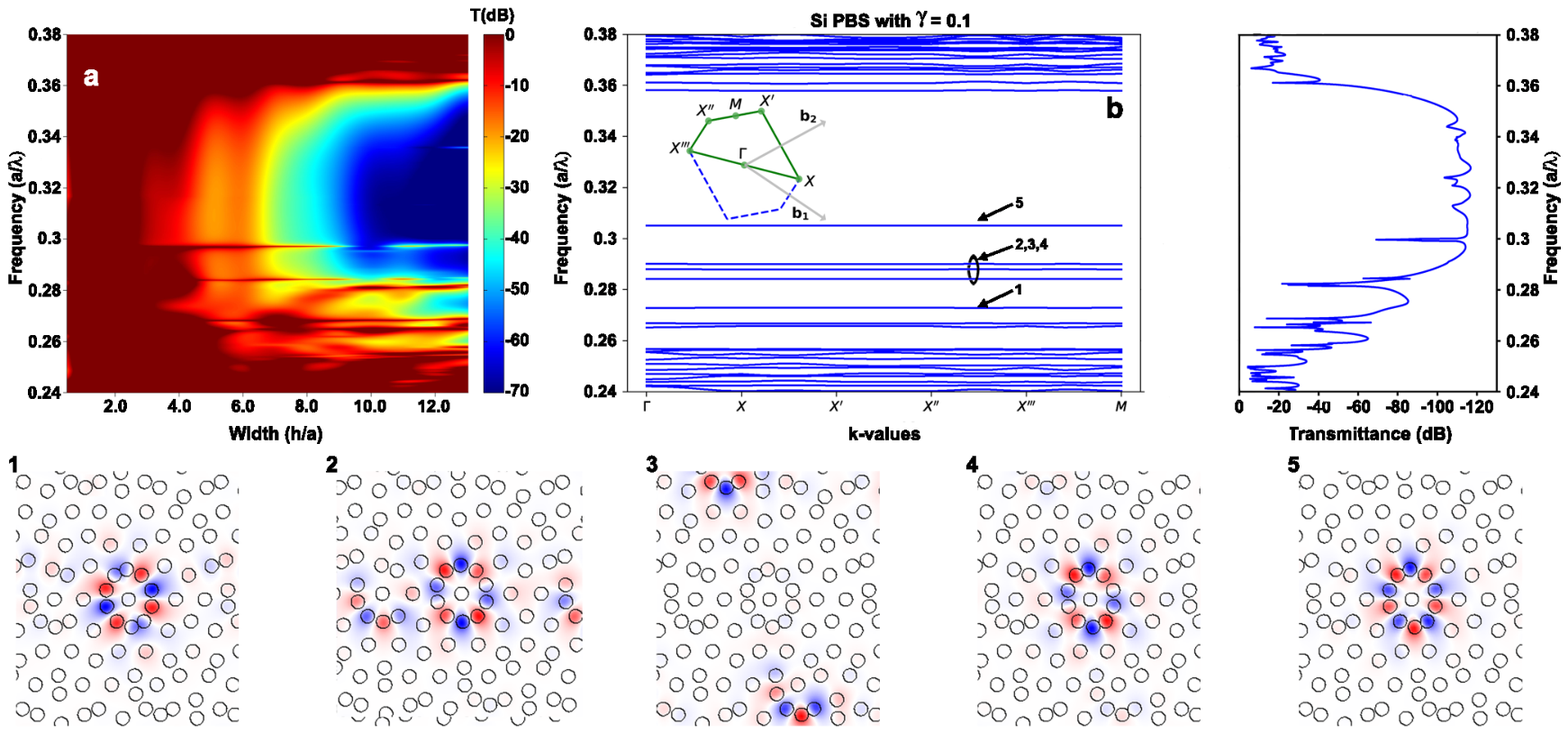}
 	\caption{\label{fig:bands_010} Sheared lattice $\gamma=0.1$. The same quantities as in Fig. \ref{fig:bands_00}.}
\end{figure*}

Results are presented in Fig. \ref{fig:bands_00}, \ref{fig:bands_005} and \ref{fig:bands_010} for different values of shear $\gamma=$0, 0.05 and 0.1 respectively. The three figures have the same structure, composed of different images that summarize all results.
 
In the top left corner (labeled by a) there is a color map that presents the transmission spectra (hotter colors higher transmission) for different widths of the sample in the horizontal axis. Frequency is presented normalized to the lattice parameter $a$ in the vertical axis  (see figure 1). Sample width, presented as $ h/a$, is also normalized to the same dimensional parameter. This representation is made by adding several spectra as the one shown in the top right figure (labeled by c) calculated with CST for different values of sample width and interpolating to get a smoother color plot. The onset of a well-defined gap as width increases can be clearly seen in this representation. In addition, several bright states inside the gap are clearly observable.

The band structure of the corresponding bulk lattice, calculated with MPB, is shown in the figures at top center (labeled by c), in the same frequency interval as the transmission data, so that comparisons between the two calculations can be readily made. The frequencies of gap edges and resonant states within the gap for the bulk structure are very close to those observed in the transmission spectra for a finite sample (simulated with CST).

We also show the electric field intensity distribution of the 5 bright states as identified in the bulk lattice band calculations (numbers 1 to 5 in the lower part of the figure). These data are calculated with MPB and presented normalized to $\pm1$ (red and blue) in all cases.


In the three cases studied, the MPB calculations of the bulk lattice and the CST simulations of finite samples are in excellent agreement. Concordance is particularly good in the first two cases, while for $\gamma=0.1$ a slight mismatch in the frequency position for state 5 can be appreciated.

Gap evolution with thickness can be seen in figures \ref{fig:bands_00}(a), \ref{fig:bands_005}(a) and \ref{fig:bands_010}(a). The general behavior is similar in the three cases. We observe that, for small thickness, attenuation is scarce at all frequencies and there is no contrast indicating a gap. A depression in transmission indicating the gap existence starts to be noticeable for widths from 3 to 6. For the central frequencies, transmission keeps diminishing while for values out of the gap values remain quite high. The maximum contrast between gap and its borders is attained for $h/a=8$.  

\begin{figure}[htbp]
	\centering
	\includegraphics[width=1\linewidth]{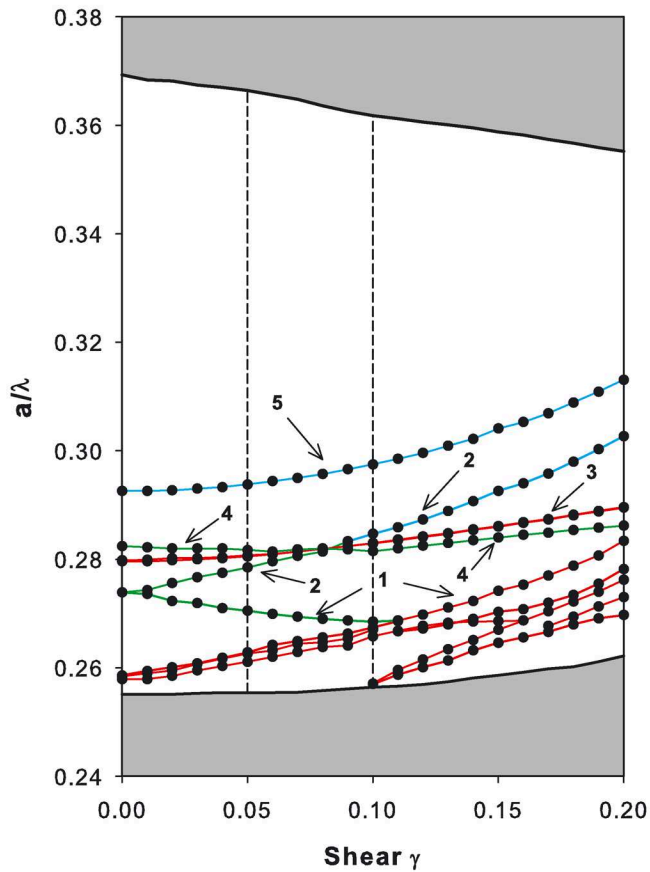}
	\caption{\label{fig:modes_shear}Frequency evolution of the resonances obtained from transmission calculations for $h/a=13$ with $\gamma$ the shear value. Solid red lines represent resonances with an average transmission intensity higher than -20 dB, solid green lines to resonances with an intensity between -20 and -40 dB and solid blue lines to resonances lower than -40 dB. Resonances are labeled from 1 to 5 as it was previously made in figures 5, 6 and 7. The edges of the band gap are represented by the grey area.Vertical dashed lines highlight the states shown in figures 6 and 7.}
\end{figure}

The gap range decreases with lattice deformation. The frequency range of low transmission is smaller as the shear increases. This happens because both gap borders move towards the center for higher values of shear. This can be seen comparing representations of transmission and band structures in figures \ref{fig:bands_00}, \ref{fig:bands_005} and \ref{fig:bands_010}. The former (\ref{fig:bands_00}(b), \ref{fig:bands_005}(b) and \ref{fig:bands_010}(b)), suggest that several bands that are packed near the gap edges for the undistorted lattice begin to open and lose degeneration when symmetry start to vanish due to deformation. In figure \ref{fig:modes_shear} we provide smooth curves of the evolution of the gap edges with shear, calculated for much more values of shear that those shown in previous figures. 

It is interesting to note that the bands inside the gap are significantly flatter than the bands present in the edge of the band gap. This characteristic shows that the light propagates very slowly at the frequency of the excited states due to the slowing down of group velocity \cite{wang2006}. This confirms that the low group velocity is essentially related to the resonant states on the local decagonal rings of the lattice and other cylinder clusters, even when shear disturbs the lattice \cite{Berman1983}.

Bright states inside the gap can be clearly seen in figures \ref{fig:bands_00}, \ref{fig:bands_005} and \ref{fig:bands_010}. The states are present in the band structure (shown as b in the three figures) and are also visible as frequencies of increased transmission in the color maps (labelled a in the figures) presenting a significant good match. These states are due to local resonances of a cylinder clusters of the structure as shown in the lower part of figures \ref{fig:bands_00} to \ref{fig:bands_005} and \ref{fig:bands_010}. The states are labeled 1 through 5 in coincidence with their naming in the band structure.

Resonance labelled 5 corresponds to a full symmetric resonance of the decagonal ring in the center of the structure, with five maxima and five minima placed in each cylinder of the ring. Resonances 1 and 2 are also held in the decagonal ring but are composed of only four maxima and minima, not fitting so well with the cylinders and allowing two configurations (90 degree from each other). These two geometrical configurations are totally equivalent and therefore their energy levels are the same, appearing degenerate in the band structure of the undistorted lattice (figure \ref{fig:bands_00}). However, this degeneration is broken when shear is introduced, and symmetry is lost.

Resonances 3 and 4 are due to incomplete decagonal rings (“C” shaped half ring), apart from the decagonal ring. Two families of incomplete rings deform differently when shear is introduced, separating their energy levels that were degenerated in the undistorted lattice. However, these two geometrical configurations remain very similar, so their energy is quite close. There are other possible resonant configurations of substructures of the lattice (close to the edge of the gap), but their energy lies outside the bandgap and cannot be considered as isolated states. 

As abovementioned, the states in the gap appear as high transmission frequencies in the spectrum (figures \ref{fig:bands_00}(a), \ref{fig:bands_005}(a) and \ref{fig:bands_010}(a)). When radiation in open space reaches the edge of a finite sample of the structure, it excites the resonance of the cylinder clusters for the appropriate frequencies. This excitation is transmitted from cluster to cluster until the other side of the sample is reached and the sample radiates again to the free space. Therefore, the radiation able to excite resonances can traverse the sample end to end metaphorically jumping in the clusters, in the same way that a person can cross a river jumping in stones placed in its course. This is the reason why the transmission at the frequency of the states remains high as the sample width is increased, while the one of other frequencies decreases much more quickly. This generates a significant contrast between the transmission of the states and its background for widths grater that $h/a$=4. A study of the optimum width to improve contrast in this kind of systems was published elsewhere \cite{andueza2016a}.    

In order to follow more closely the evolution of the states in the gap with shear, we calculated the transmission of samples of more deformation values than those presented previously as figures \ref{fig:bands_00}(c), \ref{fig:bands_005}(b) and \ref{fig:bands_010}(c). In all cases calculations are performed on finite samples wide enough to have the gap fully developed ($h/a=13$). The results are presented in figure \ref{fig:modes_shear}. 

There are many interesting features that can be derived from this figure such as the decrease  in the gap width (mainly reduced by the decreasing of the upper edge) or the increase in the number of modes that appear as deformation rises (mainly departing from the lower gap edge). 

Most of the states excited inside the band gap show a strong dependence on shear, increasing their frequency position with it. Exceptions are states 3 and 4 (that remain unaltered for certain intervals) and state 1, the only that decreases in frequency. The different slopes of these curves lead to crossings as the one observed for shear around 0.08, where states 2, 3 and 4 change their relative positions. States 1 and 2, degenerate in the undeformed system, quickly diverge in opposite directions for low values of shear. The initial frequency decrease of state 1 ends for a shear value of 0.1, and from this point on, merges with others that where rising in frequency. It is important to note that some of these states, especially states 1, 2 and 3, are quite bright, presenting a significant contrast with respect to the background, what could lead to interesting applications. We have explored elsewhere (ref. \cite{andueza2020a}) the possible construction of a strain sensor based on the frequency difference between states 1 and 2.

\section{Conclusions}

We have studied the effect of shear in the photonic band structure and transmission properties of 2D dielectric photonic quasicrystals based on cylinders with high dielectric permittivity ($\varepsilon$ = 12). The band structure of the bulk lattice (and field distribution of the resonances) was calculated with MPB, while the transmission of finite samples of different widths was calculated with CST software. The system presents an isotropic bandgap and five bright states inside the gap. These features can be observed (with a remarkable match) both, in the finite sample transmission and in the bulk photonic band structure. The photonic states in the bandgap are due to Mie like resonances of local clusters of cylinders and are associated with very flat bands, which generates very low group velocity of the transmitted radiation. The quasicrystal undergoes a remarkable electromagnetic response variation induced by shear, gap bandwidth decreases, and the distortion of the position of the cylinders, breaks energy degeneration of excited states into the bandgap and shifts the energy values of all of them. Besides, it exits a complex evolution of the frequency position of the states with shear. This set of characteristics is interesting in the design of practical devices such as sensors.

\bibliography{bibliography}

\end{document}